\documentclass[12pt]{article}
\usepackage[utf8]{inputenc}
\usepackage[left=1in,right=1in, top=1.2in,bottom=1.2in,includehead]{geometry}
\usepackage{amssymb,amsthm,amsmath,pgf}
\usepackage{mathrsfs}
\usepackage{placeins}
\usepackage{wrapfig}
\usepackage{indentfirst}
\usepackage{fancyhdr}

\usepackage{lineno}

\newtheorem{theorem}{Theorem}[section]

\newtheorem{definition}{Definition}
\newtheorem{observation}{Observation}

\begin{document}

\title{The NP-completeness of Redundant Open-Locating-Dominating Set}
\author{
    \small Robert Dohner\\
    \small Computer Science Department\\
    \small Middle Tennessee State University\\
    \small \texttt{rjd2t@mtmail.mtsu.edu }
    \and
    \small Suk J. Seo\\
    \small Computer Science Department\\
    \small Middle Tennessee State University\\
    \small \texttt{Suk.Seo@mtsu.edu}
}
\date{}

\maketitle
\thispagestyle{empty}

\begin{abstract}
For a graph $G$, a dominating set $D$ is a subset of vertices in $G$ where each of the vertices in $G$ is  in $D$ or adjacent to some vertex in $D$. An open-locating-dominating (OLD) set models a system with sensors to detect an intruder in a facility or a faulty component in a network of processors. The goal is to detect and pinpoint an intruder's exact location in a system with a minimum number of sensors.  In this paper, we focus on a fault-tolerant OLD set called a redundant OLD set and present a proof for the NP-completeness of the problem of a redundant OLD set.
\end{abstract}

\noindent
\textbf{Keywords:} \textit{NP-completeness, domination, distinguishing set, \\
redundant open-locating-dominating set}
\vspace{1em}


\section{Introduction} 
Let $G = (V(G), E(G))$ denote an undirected graph with its vertex set $V(G)$ and its edge set $E(G)$.
For a pair of vertices $u$ and $v$, if $uv \in E(G)$, then $u$ and $v$ are said to be adjacent. The \textit{open neighborhood} of a vertex $v$, denoted by $N(v)$, is
the set of vertices adjacent to $v$: $N(v)=\{w\in V(G): vw\in E(G)\}$, and the \textit{closed neighborhood} of $v$ is $N[v]=N(v)\cup \{v\}$.  Vertex set $D \subseteq V(G)$ is an \textit{open dominating set} (also called a \textit{total dominating set}) if  $\cup _{x \in D} N(x) = V(G)$.

A graph is used to model a system of setting up sensors to detect a thief in a facility or a malfunctioning computing element in a network of processors, where each vertex represents a possible location of an ``intruder" or a malfunctioning computing element. For detection, various types of subsets of vertices are used with the following two properties:  one is to cover the entire set of vertices in the system, and the other is to uniquely identify a vertex with an intruder.  These sets implicitly form \textit{distinguishing sets} and they include an identifying code \cite{charon, karpovsky}, a locating-dominating set \cite{honkala, slater1987} and an open-locating-dominating set \cite{chellali, seo2011}.  There is a bibliography of currently over 470 papers on the topic of distinguishing sets that are maintained by Lobstein \cite{lobstein}.

In this paper, we consider a fault-tolerant open-locating-dominating (OLD) set called a redundant open-locating-dominating (RED:OLD) set.  In general, a fault-tolerant distinguishing set maintains its properties even when the detection device at a vertex is faulty (See  \cite{seo2015, seo2018_2,slater2002} for details.).  

\begin{definition}
 An open dominating set $S \subset V(G)$ is an \textit{open-locating-dominating set} if for all $u$ and $v$ in $V(G)$ one has $N(u) \cap S \neq N(v) \cap S$. 
\end{definition}

Clearly, a graph $G$ has an open-locating-dominating set only when no two vertices have the same open neighborhood.  Next, we present a theorem that characterizes an OLD set, but first, we describe the concept of \textit{k-distinguishing} in the context of an open-dominating set.

\begin{definition}
 A pair of vertices, $u$ and $v$ in $V(G)$ is said to be \textit{k}-\textit{distinguished} by a subset $S\subset V(G)$ if $ | (N(u) \cap S)  \triangle (N(v) \cap S)| \geq k$, where the symbol $\triangle$ denotes the symmetric difference.
\end{definition}

In an open-locating-dominating set $S$ of a graph $G$, for a pair of vertices $u$ and $v$ in $V(G)$, there must be 1) a vertex in $S$ that dominates $u$, but not $v$ or 2) a vertex that dominates $v$, but not $u$. In other words, vertices $u$ and $v$ must be 1-distinguished.

\begin{theorem}
A subset $S \subseteq V(G)$ for a $G$ is an open-locating-dominating set if and only if 1) each vertex $v \in V(G)$ is open-dominated at least once in S and 2) each pair $u$, $v$ of distinct vertices in $V(G)$ is  1-distinguished by S.
\label{theorem_old}
\end{theorem}

\begin{definition}
A collection   $C= \{S_1, S_2, ..., S_p\}$ of subsets of $V(G)$ is a \textit{distinguishing set} for a graph \textit{G} if $ \cup_{1\leq i \leq p} S_i = V(G)$ and for every pair of distinct vertices \textit{u} and \textit{v} in $V(G)$ some $S_i$ contains exactly one of them. 
\end{definition}

Using Definition 3, we can describe the aforementioned three distinguishing sets as follows. A subset of $V(G)$, $D_1 = \{w_1, w_2,..., w_j\}$ is a \textit{locating-dominating set} if $C_1 = \{\{w_1\}, N(w_1), \{w_2\},$ $N(w_2),..., \{w_j\}, N(w_j)\}$ is distinguishing; $D_2=\{w_1, w_2,..., w_j\}$ is an \textit{identifying code} if $C_2= \{N[w_1], N[w_2],...,N[w_j]\}$ is distinguishing; and $D_3=\{w_1, w_2, . . ., w_j\}$ is an \textit{open-locating-dominating set} if $C_3= \{N(w_1), N(w_2),...,$ $N(w_j)\}$ is distinguishing.

A redundant OLD-set (redundant distinguishing set) for $G$ is an OLD-set (distinguishing set) where it fully functions even when we have a malfunctioning detector in the system, which is modeled by $G$.

\begin{definition}
An open-dominating set $S \subseteq V(G)$ is called a \emph{redundant open-locating-dominating (RED:OLD)} set if $\forall v \in S$, $S - \{v\}$ is an OLD set.
\end{definition}

We observe that in a RED:OLD set every vertex has to be dominated at least twice. The following theorem characterizes a RED:OLD set, and its proof is given in Seo and Slater \cite{seo2015}. 

\begin{theorem}
A subset $S \subseteq V(G)$ for a G is a redundant open-locating-dominating set if and
only if 1) each vertex $v \in V(G)$ is open-dominated at least twice in S and 2) each pair $u$, $v$ of distinct vertices in $V(G)$ is  2-distinguished by S.
\label{theorem_red}
\end{theorem}

A trivial method of forming an OLD-set or a RED:OLD set for a graph $G$ would be using every vertex in $V(G)$, given that those sets are defined. Naturally, we are interested in finding the smallest possible size for both an OLD-set and a RED:OLD set.  We use OLD(G) and RED:OLD(G) to denote the minimum cardinality of an OLD-set and a RED:OLD set, respectively. If the cardinality of an OLD-set for $G$ achieves the value of OLD(G), then it is called an OLD(G)-set.  Similarly, a RED:OLD set has the cardinality equal to the value of RED:OLD(G), then it is called a RED:OLD(G)-set. 

\begin{figure}[ht]
    \centering
    \begin{tabular}{cc}
        \begin{tabular}{c}\includegraphics[width=0.41\textwidth]{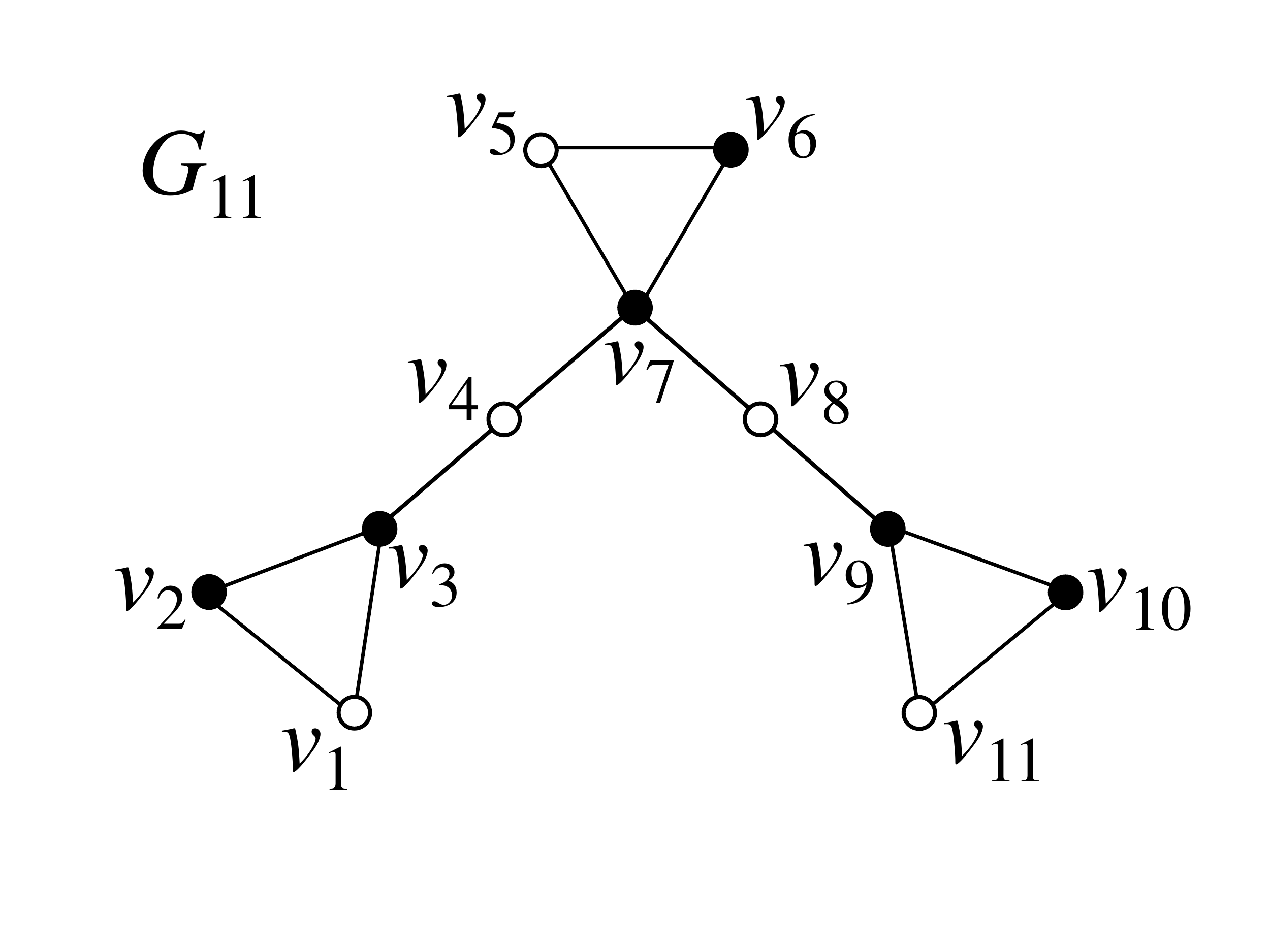}\\(a)\end{tabular} &
        \begin{tabular}{c}\includegraphics[width=0.41\textwidth]{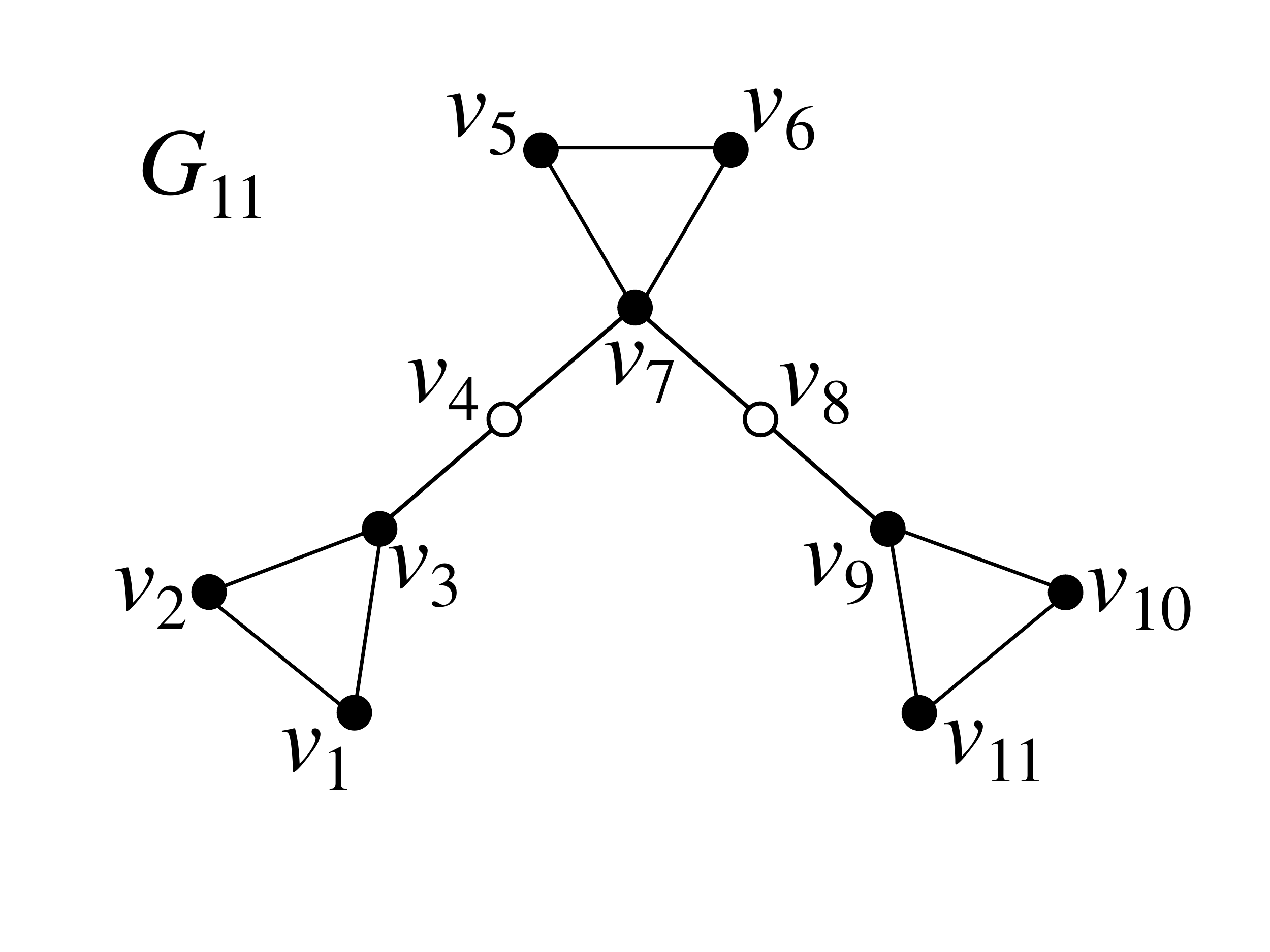}\\(b)\end{tabular}
    \end{tabular}
    \caption{(a) OLD($G_{11}$) = 6 (b) RED:OLD($G_{11}$) = 9}
    \label{fig:redold}
\end{figure}

\begin{observation}
For the graph $G_{11}$ in Figure \ref{fig:redold}, $OLD(G_{11})=6$.
\end{observation}
\noindent\begin{proof}
Consider the vertex subset $S=\{v_2,v_{3},v_{6},v_{7},v_{9},v_{10}\}$ shown as shaded in Figure \ref{fig:redold}. Clearly, $S$ is an open-dominating set for $G_{11}$ and we can verify that for every vertex pair, $u$ and $w$, we have $N(u) \cap S \neq N(w) \cap S$. For example, we have $S \cap N(v_1) = \{v_2, v_3\}$, $S \cap N(v_2) = \{v_3\}$, and $S \cap N(v_3) = \{v_2\}$, so all vertex pairs in $\{v_1, v_2, v_3\}$ are distinguished.  By symmetry, all vertex pairs in $\{v_5, v_6, v_7\}$ and in $\{v_9, v_{10},v_{11}\}$ are also distinguished.  Finally, we have $S \cap N(v_4) = \{v_3, v_7\}$, $S \cap N(v_8) = \{v_7, v_9\}$.  Hence, by Definition 1, $S$ is an open-locating-dominating set.  Note that $S$ also meets the two requirements given in Theorem \ref{theorem_old}.  Further, we can consider a collection of subsets, $N(w)$ for each $w \in S$, as described in Definition 3.  Then, for $S$ we obtain $C= \{N(v_2), N(v_3), N(v_6), N(v_7), N(v_9), N(v_{10})\}$, where $N(v_2)$ is $\{v_1,v_3\}$, $N(v_3)$ is $\{v_1,v_3\}$, and so on. For every pair of distinct vertices \textit{u} and \textit{v} in $G_{11}$ some member in $C$ contains exactly one of them. 
 Finally, we can verify that there are no other OLD-sets whose cardinality is less than $|S|=6$, and hence $S$ is an OLD($G_{11}$)-set and OLD($G_{11}$) = 6.  
\end{proof}

\begin{observation}
For the graph $G_{11}$ in Figure \ref{fig:redold}, $RED$:$OLD(G_{11})=9$.
\end{observation}
\noindent\begin{proof}
The set $S=\{v_1,v_{2},v_{3},v_5,v_{6},v_{7},v_{9},v_{10},v_{11}\}$ is a RED:OLD-set for $G_{11}$. Let $R_i = S -\{v_i\}$ for $v_i \in S$. We can verify that each $R_i$ is an OLD-set.  Hence, $RED$:$OLD(G_{11}) \leq 9$.  
Let $T$ be any RED:OLD set for $G_{11}$. Consider the vertex set $\{v_1, v_2, v_3\}$, which forms a triangle.  By Theorem \ref{theorem_red}, every vertex in $V(G_{11})$ has to be double-dominated and we observe that $R_1$, $R_2$, or $R_3$ is not a RED:OLD set, so we must have $\{v_1,v_{2},v_{3}\} \subset T$. By symmetry, we must have  $\{v_5, v_6, v_7\} \subset T$ and $\{v_{9},v_{10},v_{11}\} \subset T$. Hence, $RED$:$OLD(G_{11}) \geq 9$ and $S$ is the unique $RED$:$OLD(G_{11})$-set.
\end{proof}

In the next section, we consider RED:OLD sets for various infinite graphs, and in Section 3, we present a proof for the NP-completeness of the problem of determining RED:OLD(G).

\section{Redundant OLD-set for infinite graphs}

Much work has been done involving OLD-set parameters for various graphs. In this section, we will look at RED:OLD sets for four popular infinite grids, namely the square, the hexagonal, the triangular, and the king grids. Because each of these is an infinite graph, we use the notation RED:OLD\%(G) instead of RED:OLD(G).  For an infinite graph $G$, \textit{RED:OLD\%(G)} denotes the minimum possible value of the density of a RED:OLD set, which is measured by the percentage of the vertices in the RED:OLD set over the entire vertices in $G$.

\begin{figure}[ht]
    \centering
    \begin{tabular}{cc}
        \begin{tabular}{c}\includegraphics[width=0.45\textwidth]{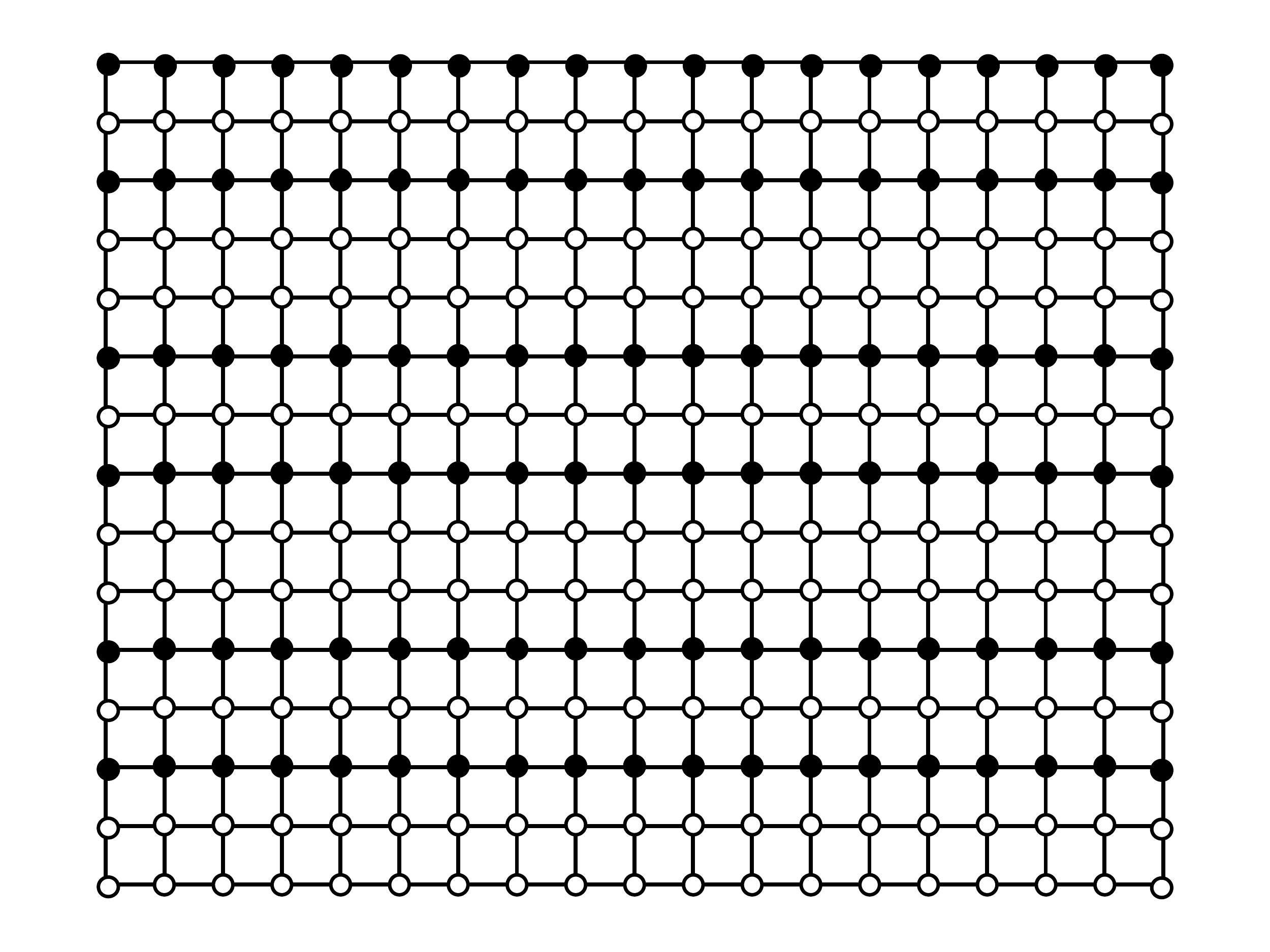}\\(a)\end{tabular} &
        \begin{tabular}{c}\includegraphics[width=0.45\textwidth]{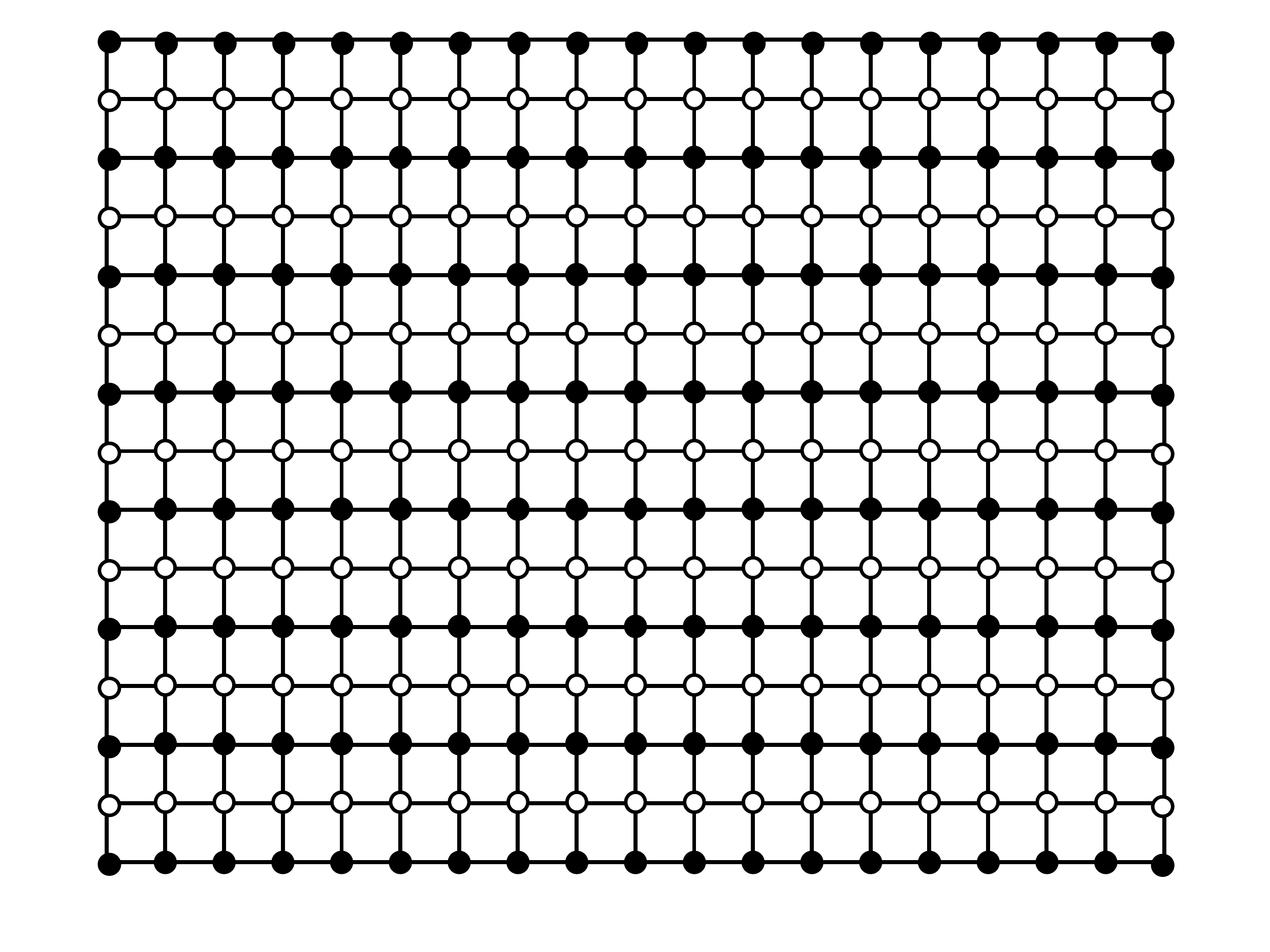}\\(b)\end{tabular}
    \end{tabular}
    \caption{OLD\%(SQ) $\le 2/5$ and RED:OLD\%(SQ) $\le 1/2$}
    \label{fig:sq}
\end{figure}

The infinite square grid, denoted as SQ, is a 4-regular graph, that is, every vertex is of degree four.  Figure \ref{fig:sq} shows an OLD set and a RED:OLD set for the SQ graph with densities 2/5 and 1/2, respectively, which establish upper bounds of the minimum density. Hence, we have OLD\%(SQ) $\le 2/5$ and RED:OLD\%(SQ) $\le 1/2$ and both of the upper bounds have also been proven to be theoretical lower bounds.

\begin{theorem}
For the infinite square grid SQ, \newline
(a) (Seo and Slater \cite{seo2010}) OLD\%(SQ) = 2/5. \newline
(b) (Seo and Slater \cite{seo2015}) RED:OLD\%(SQ) = 1/2.
\end{theorem}

\begin{figure}[ht]
    \centering
    \begin{tabular}{cc}
        \begin{tabular}{c}\includegraphics[width=0.45\textwidth]{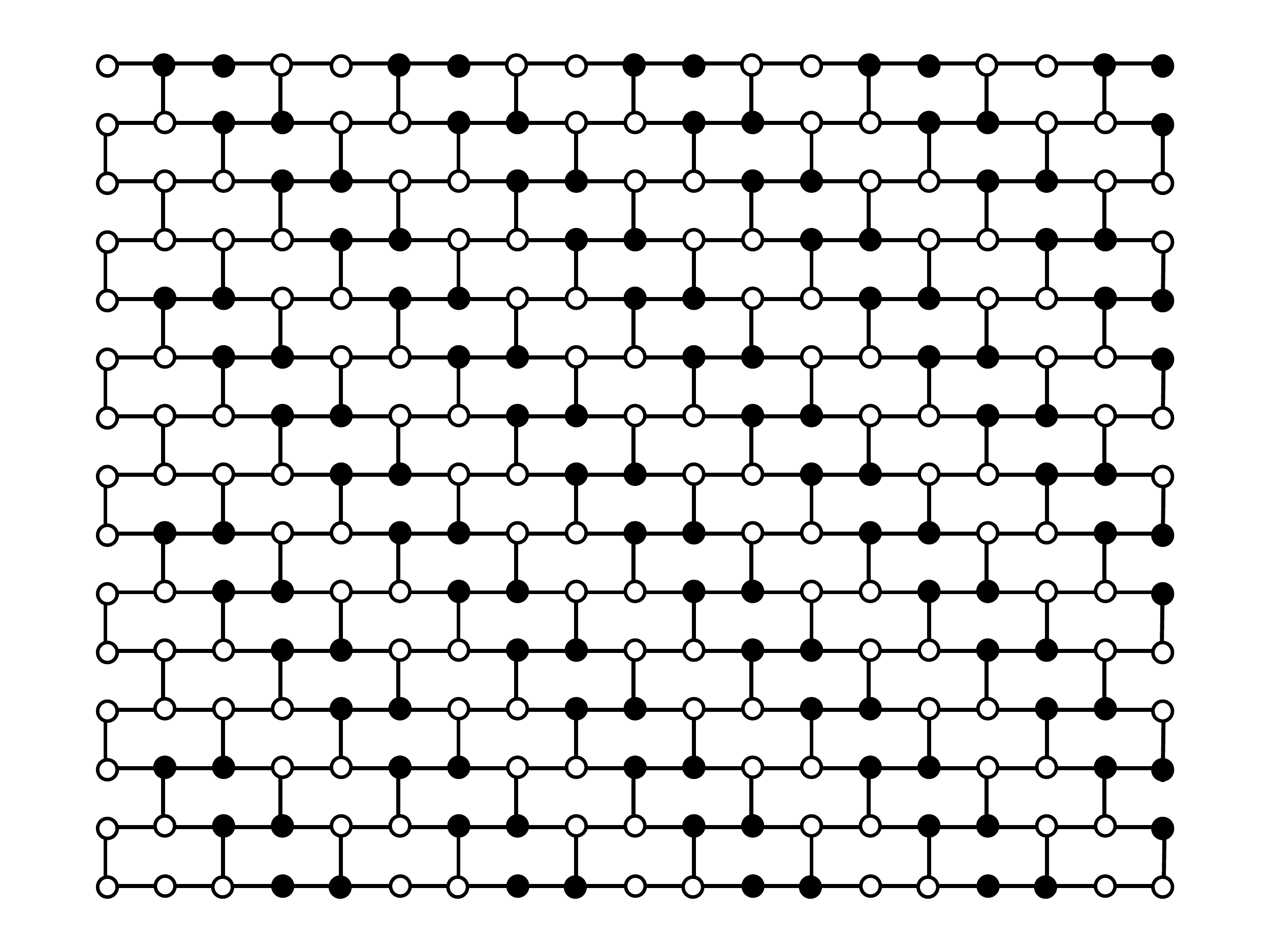}\\(a)\end{tabular} &
        \begin{tabular}{c}\includegraphics[width=0.45\textwidth]{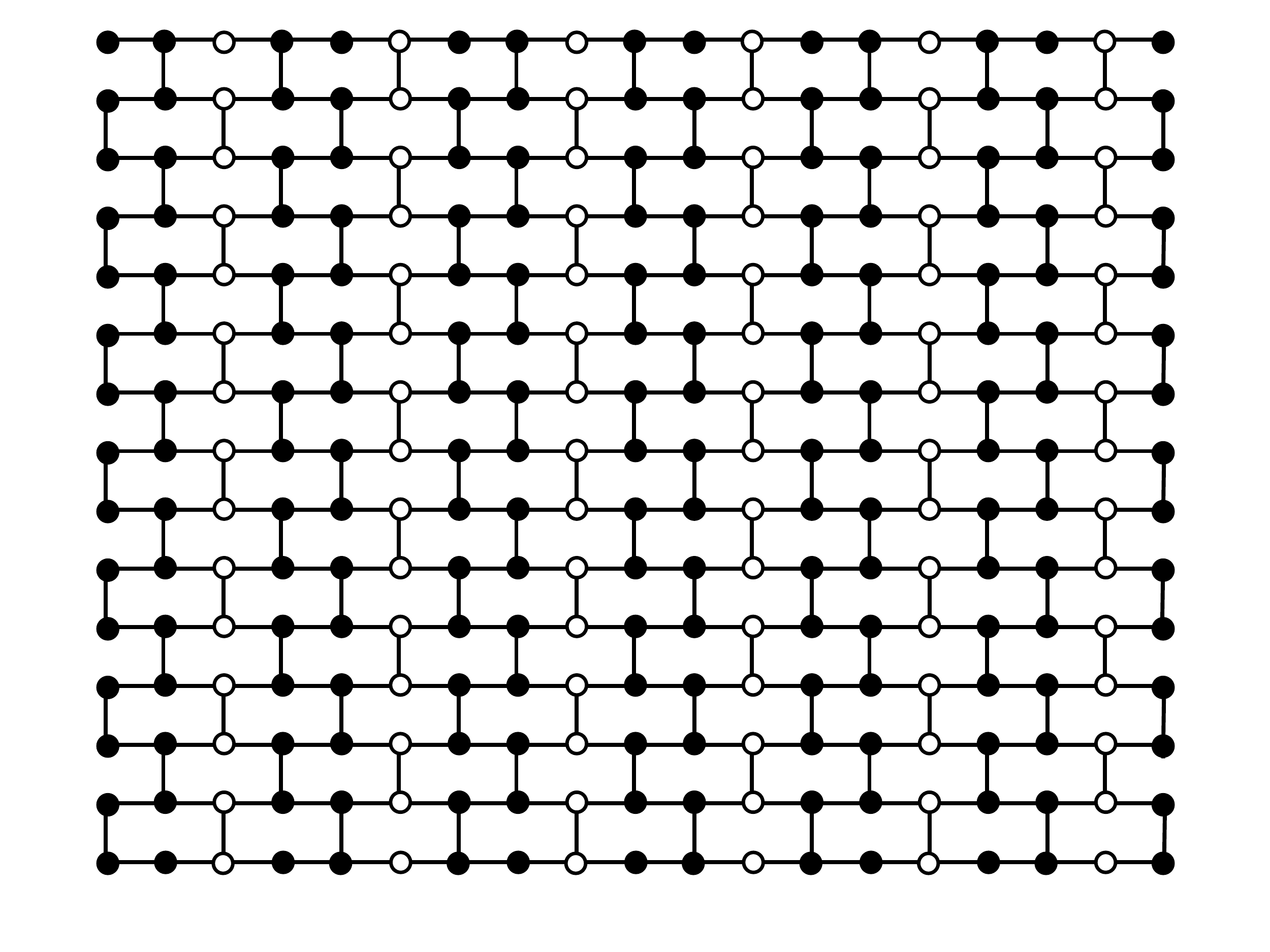}\\(b)\end{tabular}
    \end{tabular}
    \caption{OLD\%(HEX) $\le 1/2$ and RED:OLD\%(HEX) $\le 2/3$}
    \label{fig:hex}
\end{figure}

The infinite hexagonal grid, denoted as HEX, is a 3-regular graph.
Figure \ref{fig:hex} shows an OLD set and a RED:OLD set for the HEX graph with densities 1/2 and 2/3, respectively, and hence we have OLD\%(HEX) $\le 1/2$ and RED:OLD\%(HEX) $\le 2/3$.  These have also been proven to be theoretical lower bounds.

\begin{theorem}
For the infinite hexagonal grid HEX, \newline
(a) (Seo and Slater \cite{seo2010}) OLD\%(HEX) = 1/2. \newline
(b) (Seo and Slater \cite{seo2015}) RED:OLD\%(HEX) = 2/3.
\end{theorem}

\begin{figure}[ht]
    \centering
    \begin{tabular}{cc}
        \begin{tabular}{c}\includegraphics[width=0.45\textwidth]{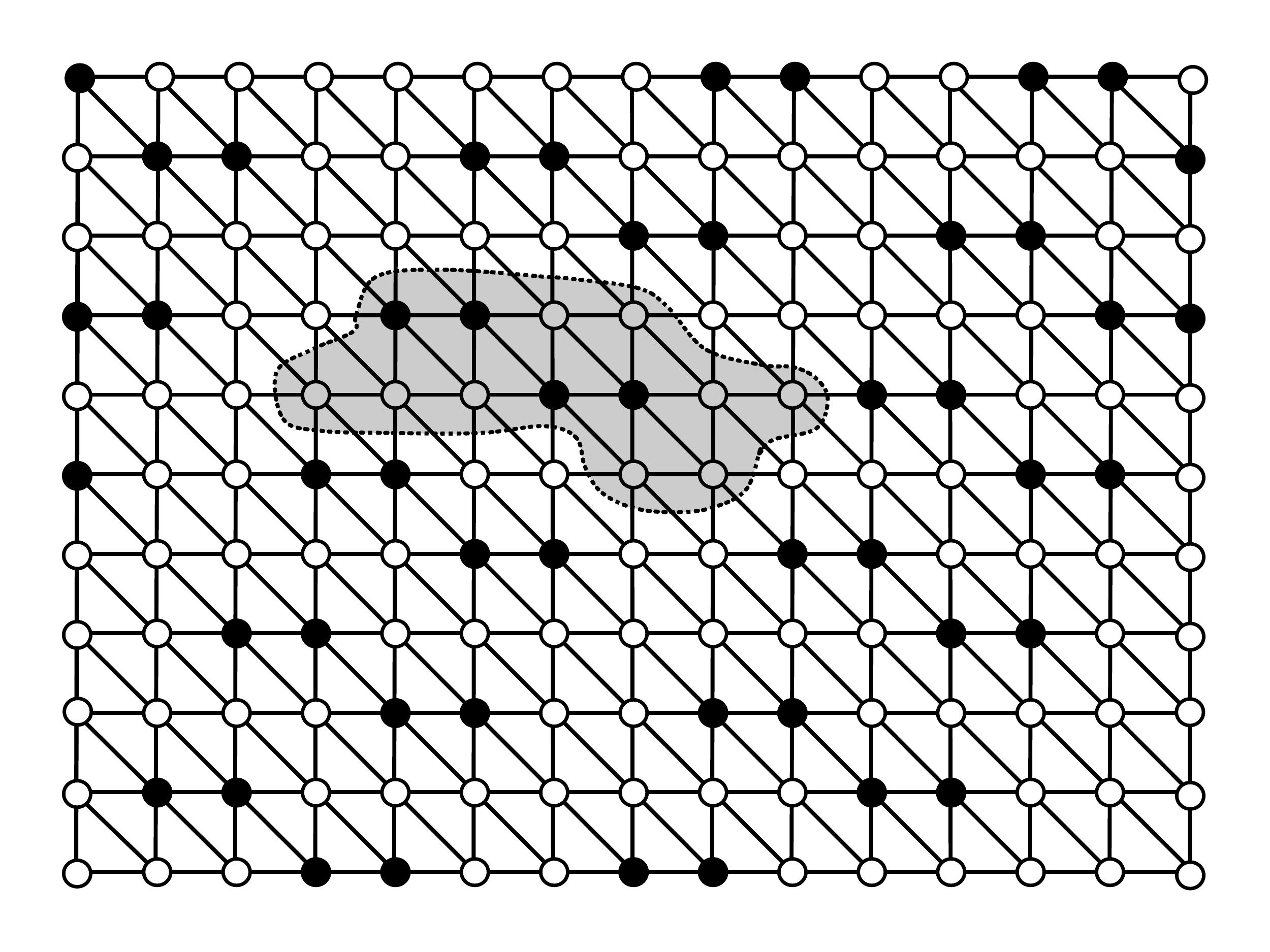}\\(a)\end{tabular} &
        \begin{tabular}{c}\includegraphics[width=0.45\textwidth]{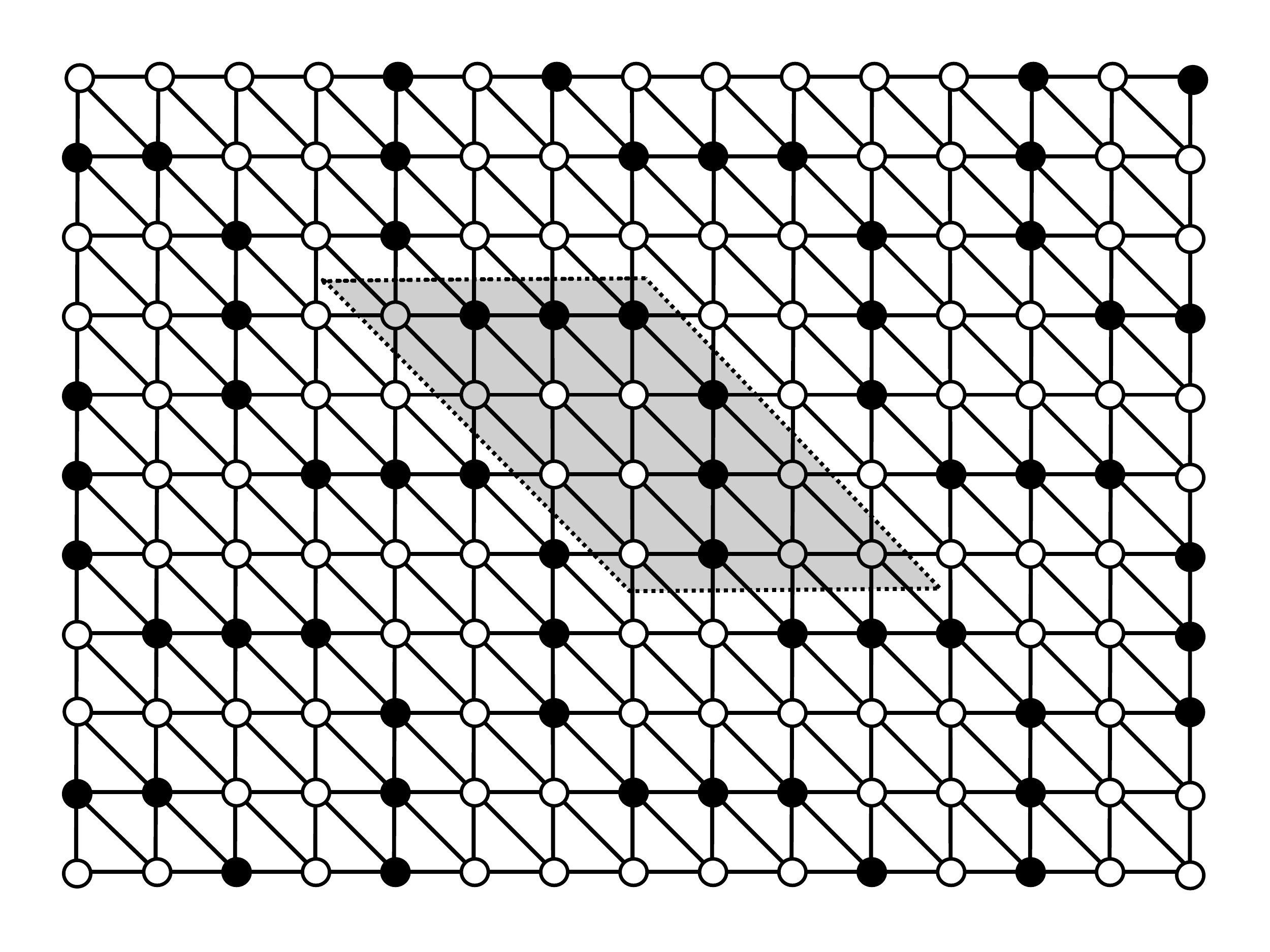}\\(b)\end{tabular}
    \end{tabular}
    \caption{OLD\%(TRI) $\le 4/13$ and RED:OLD\%(TRI) $\le 3/8$}
    \label{fig:tri}
\end{figure}

The infinite triangular grid, denoted as TRI, is a 6-regular graph. Seo and Slater \cite{seo2010} initially showed that OLD\%(TRI) is at most 1/3.  However, later Kincaid et al. \cite{kincaid} not only improved the upper bound to be 4/13 
, but also proved that 4/13 is a theoretical lower bound confirming Theorem \ref{theo_old_tri} shown below. Note that the ``amoeba" shape of the tile, as shown in Figure \ref{fig:tri} (a) has the density of 4/13, and its tiling can cover the entire graph. Hence, this solution establishes that OLD\%(TRI) $\le 4/13$. 
Figure \ref{fig:tri} (b) shows a RED:OLD set with a parallelogram shape of tile with density 3/8, which was discovered by Seo and Slater \cite{seo2015}.  Again, this tiling can cover the entire graph, and hence establishing an upper bound of 3/8: RED:OLD\%(TRI) $\le 3/8$.  They proved that the solution with the density of 3/8  is optimal, as shown in the next theorem.

\begin{theorem}
For the infinite triangular grid TRI, \newline
(a) (Kincaid et al. \cite{kincaid}) OLD\%(TRI) = 4/13. \newline
(b) (Seo and Slater \cite{seo2015}) RED:OLD\%(TRI) = 3/8.
\label{theo_old_tri}
\end{theorem}

Now we consider the infinite king grid, denoted as KING, which is 8-regular.
Seo \cite{seo2018_2} looked into OLD sets in the infinite king’s graph and discovered that there are multiple OLD sets, all of which achieve a density of 1/4. Using the “open-share” argument, Seo was able to find the theoretical lower bound for OLD sets in the infinite king’s graph as 6/25, which is slightly less than 1/4. We have discovered RED:OLD sets with density 1/3, which is the best upper bound to date, and there are no known non-trivial theoretical lower bounds.

\begin{theorem}
For the infinite king grid KING, \newline
(a) (Seo \cite{seo2018}) 6/25 $\le$ OLD\%(KING) $\le$ 1/4. \newline
(b) RED:OLD\%(KING) $\le$ 1/3.
\label{theorem_king}
\end{theorem}

\section{NP-completeness of the redundant OLD set problem} 

The problem of finding an OLD(G) for an arbitrary graph $G$ is known to be NP-complete \cite{seo2010}, and here we present a proof for the NP-completeness of the problem of determining $RED$:$OLD(G)$, the least cardinality of a RED:OLD set for $G$.  See Garey and Johnson \cite{garey} for details on NP-completeness.  
We will show the proof using the reduction from the 3-SAT problem. \newline

\noindent
\textbf{3-SAT} \newline
\textbf{INSTANCE:}  Sets C = \{$c_1$, $c_2$, \ldots,$c_M$\} of clauses on U = \{$u_1$, $u_2$, \ldots, $u_N$\} of variables such that $|$$c_i$$|$ = 3 for 1 $\leqslant$ i $\leqslant$ M. \newline
\textbf{QUESTION:}  Is there a satisfying truth assignment for C? \newline

\noindent
\textbf{RED:OLD (Redundant open-locating-dominating set problem)} \newline
\textbf{INSTANCE:}  Graph G = (V, E). \newline
\textbf{QUESTION:}  Does G have a RED:OLD set with RED:OLD(G) $\leqslant$ K?

\begin{figure}[ht]
    \centering
    \begin{tabular}{cc}
        \begin{tabular}{c}\includegraphics[width=0.35\textwidth]{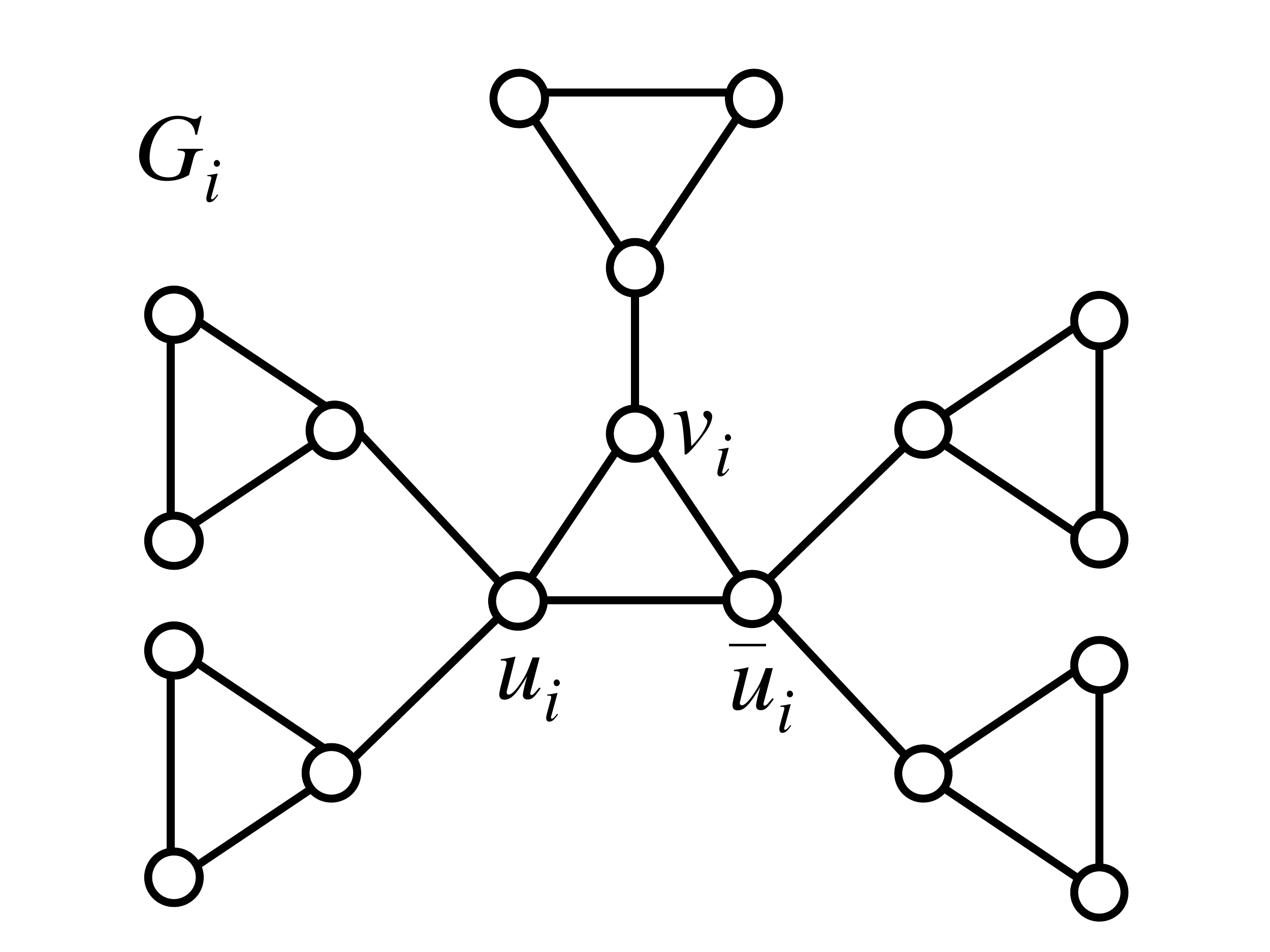}\\(a)\end{tabular} &
        \begin{tabular}{c}\includegraphics[width=0.35\textwidth]{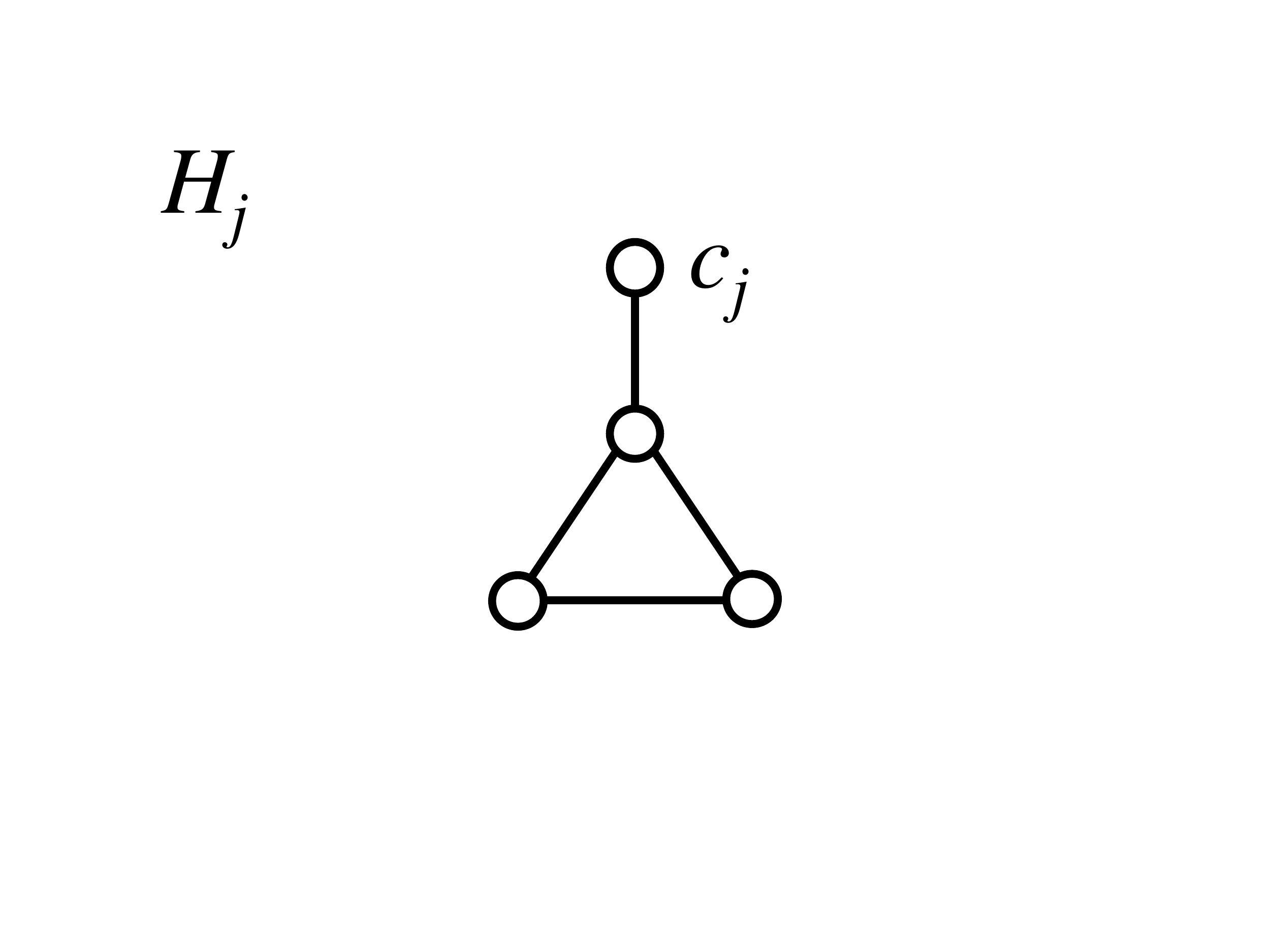}\\(b)\end{tabular}
    \end{tabular}
    \caption{(a) a variable graph  (b) a clause graph }
    \label{fig:variable}
\end{figure}

\begin{theorem}
Problem RED:OLD set is NP-complete.
\end{theorem}

\begin{figure}[ht]
    \centering
    \includegraphics[width=0.8\textwidth]{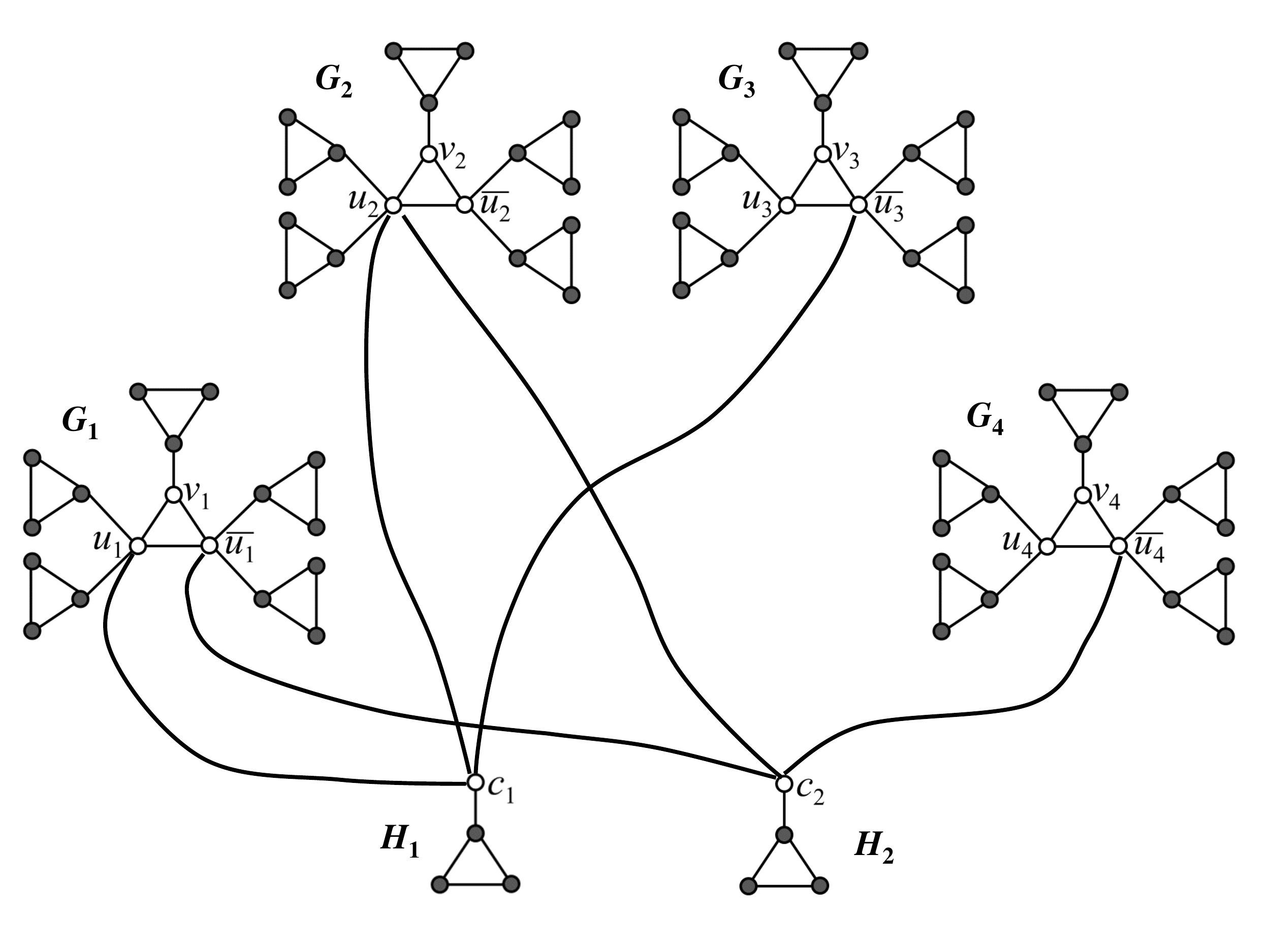}
    \caption{Construction of $G$ from $c_1 = \{u_1, u_2, \overline{u_3}\}$ and $c_2 = \{\overline{u_1}$, $u_2$, $\overline{u_4}\}$}
    \label{fig:exampleclause}
\end{figure}

\begin{proof}
First, Problem RED:OLD belongs to class NP because it can be solved by a non-deterministic polynomial-time algorithm and a candidate solution can be verified in deterministic polynomial-time.  We will show a polynomial-time reduction from 3-SAT to RED:OLD.

For each $u_i \in U$, construct the graph $G_i$ on 18 vertices as shown in Figure \ref{fig:variable}.  For each clause $c_j \in C$, construct the graph $H_j$ on 4 vertices, also shown in Figure \ref{fig:variable}.  To complete the creation of the graph G, for $C_j$ with 1 $\leqslant$ j $\leqslant$ M if clause $C_j$ is \{$u_{j,1}$, $u_{j,2}$, $u_{j,1}$\}, where each $u_{j,t}$ is some $u_i$ or  $\overline{u_i}$, let clause vertex $c_j$ be adjacent to variable vertices $u_{j,1}$, $u_{j,2}$, and $u_{j,3}$.  Figure \ref{fig:exampleclause} shows a complete construction of $G$ from two clauses $c_1 = \{u_1$, $u_2$, $\overline{u_3}$\} and $c_2 = \{\overline{u_1}$, $u_2$, $\overline{u_4}$\}.  We see that $G$ has 18N + 4M vertices and can be constructed from $C$ in polynomial-time.

Assume that $S$, a subset of $V(G)$, is a RED:OLD(G)-set.  Then, S is required to contain every shaded vertex in $G_i$, as shown in Figure \ref{fig:exampleclause}.  Additionally, for vertex $v_i$ in each $G_i$ to be double dominated, either $u_i$ or $\overline{u_i}$ must be a member of the set S.  For each $H_j$, the three vertices that form a triangle must be in $S$ as shown in Figure \ref{fig:exampleclause}. Now $S$ contains 16N+3M vertices.  Since $S$ is a RED:OLD set, vertex $c_j$ in each $H_j$ must be dominated by at least one vertex from the set of three literal vertices, implying there is a satisfying truth assignment for clause $c_i$.

Suppose there is a satisfying truth assignment of $C$.  We first form a set $S$ containing all of the shaded vertices in Figure \ref{fig:exampleclause}.  Now for each literal, we add $u_i$ to $S$ if it is true.  Otherwise, we add $\overline{u_i}$.  We observe that each $c_i$ vertex will be dominated by at least one other vertex from the set of three literal vertices because there is a satisfying truth assignment of $C$. We see that $S$ is a RED:OLD(G) set containing 16N+3M vertices.

We have shown the reduction of 3-SAT to the RED:OLD problem, and it can be done in polynomial-time.  We conclude that RED:OLD(G) = 16N+3M if and only if there is a satisfying truth assignment of $C$.
\end{proof}

\end{document}